\documentclass[referee]{aastex}
\usepackage[usenames]{color}
\usepackage{epstopdf}
\usepackage{txfonts}
\usepackage{color}

\shorttitle{Infrared and optical polarimetry around IRAS 4A}
\shortauthors{F.~O.~Alves et al.}

\newcommand{\kms}{km~s$^{-1}$}

\begin{document}

\title{Infrared and optical polarimetry around the low-mass star-forming region
NGC\,1333  IRAS\,4A\slugcomment{Based on observations collected  with the 4.2 m
William Herschel Telescope, at La Palma, Canary Islands (Spain)  and the 1.6 m
Telescope of the Observat\'orio do Pico dos Dias, operated by  Laborat\'orio
Nacional de Astrof\'\i sica (LNA/MCT, Brazil) .}}

\author{F. O. Alves$^1$, J. A. Acosta-Pulido$^{2,3}$, J. M. Girart$^1$, G. A. P.
Franco$^4$, and R. L\'opez$^5$ }

\affil{
$^1$Institut de Ci\`encies de l'Espai (IEEC--CSIC), Campus UAB,
	Facultat de Ci\`encies, C5 par 2$^{{\mathrm a}}$, 08193 Bellaterra,
	Catalunya, Spain\\
	\email{[oliveira;girart]@ice.cat}
$^2$Instituto de Astrof\'isica de Canarias, V\'ia L\'actea s/n, E-38200 La
	Laguna, Tenerife,  Spain \\ \email{jap@iac.es}	    
$^3$Departamento de Astrof{\'i}sica, Universidad de La Laguna, E-38205 La
	Laguna, Tenerife, Spain \\
$^4$Departamento de F\'isica -- ICEx -- UFMG, Caixa Postal 702, 30.123-970 Belo 
	Horizonte, Brazil\\
	\email{franco@fisica.ufmg.br}
$^5$Departament d'Astronomia i Meteorologia (IEEC-UB), Institut de Ci\`encies
	del Cosmos, Universitat de Barcelona, Mart\'{\i} i Franqu\`es 1,
	E-08028 Barcelona, Spain \\
	\email{rosario@am.ub.es}
}

\begin{abstract}
We performed $J$- and $R$-band linear polarimetry with the 
4.2~m William Herschel Telescope at the Observatorio del Roque de los Muchachos
and with the 1.6~m telescope
at the Observat\'orio do Pico dos Dias, respectively, to derive the magnetic
field geometry of the diffuse molecular cloud 
surrounding the embedded  protostellar 
system NGC\,1333 IRAS\,4A. We obtained 
interstellar polarization data for 
about two dozen stars. The distribution of polarization position angles has low
dispersion  and suggests the existence of an ordered magnetic field component 
at physical scales larger than the protostar. 
Some of the  observed stars present intrinsic polarization and evidence of
being young stellar  objects. The estimated mean orientation of the
interstellar magnetic field as derived from  these data is almost perpendicular
to the main direction of the magnetic field associated  with the dense
molecular envelope around IRAS\,4A. Since the distribution of the CO  emission
in NGC 1333 indicates that the diffuse molecular gas has a multi-layered 
structure, we suggest that the observed polarization position angles are caused
by the  superposed projection along the line of sight of different magnetic
field components.
\end{abstract}


\section{Introduction}
 
Infrared and optical polarimetry is a suitable tool for observing magnetic fields
within molecular clouds at large scales. At these wavelengths polarization 
can be produced by dichroic extinction of background 
starlight. \citet{Davis51} proposed that a fraction of
non-spherical interstellar dust grains become aligned perpendicular to the
local magnetic field due to paramagnetic relaxation. Although this mechanism is 
commonly invoked in the literature, it seems to be inefficient within molecular clouds 
\citep[e.g.,][]{Lazarian07}. However, a more realistic scenario was proposed by 
several authors who have successfully modeled the perpendicular alignment 
between grains and
magnetic fields by radiative torques propelled by anisotropic radiation 
\citep{Draine96,Lazarian07b,Hoang08,Hoang09}. 
 
Aligned dust grains behave like a polarizer to any incoming radiation,
absorbing and scattering the component of the electric field ($E$-vectors)
parallel to their longest axis. Therefore,  the observed radiation will carry
some degree of linear polarization. 
The resulting polarization map outlines the geometry of the magnetic field
lines projected  onto the plane of sky (POS).
Near-infrared (near-IR) polarimetric 
observations trace visual extinctions of a few tens of magnitudes, providing
deeper photometry than optical wavelengths. However,  the increase in
interstellar extinction is not usually accompanied by a 
linear increase in the degree of polarization. 
This has been interpreted as a decrease in the polarization efficiency, or
depolarization, with increasing  visual extinction
\citep{Goodman92,Goodman95,Gerakines95,Arce98}.
Nevertheless, this depolarization at optical wavelengths is not observed in the
Pipe Nebula \citep{Franco10} and submillimeter polarization observations show
that there is unequivocal evidence that grains do align in dense environments
with high visual extinction \citep[e.g.,][]{Whittet08,Vaillancourt08}. 
Additionally, 
the scattering of stellar light by dust grains also generates linear polarization
in the optical and near-IR. This type of polarization is found in
reflection nebulae  associated with disks and envelopes of  young stars.
 
NGC 1333 is the most active star-forming site in the Perseus molecular cloud 
\citep{Lada96}. 
A large portion of the NGC~1333 young stellar cluster is composed by low-mass
stars younger than 1 Myr  \citep{Wilking04}. In addition, there are numerous
embedded protostars powering molecular and Herbig-Haro outflows \citep{Knee00}.
There is evidence that the molecular cloud in NGC 1333 is being disturbed by
the large amount of outflow  \citep{Warin96,Sandell01,Quillen05}.

The first polarimetric observations  toward NGC\,1333 were carried out by
\citet{Vrba76} and \citet{TTC80}. \citet{Tamura88} conducted $K$-band
polarimetric observations towards the center of the NGC 1333 reflection nebula.
A larger polarimetric survey covering the full Perseus complex was carried out by
\citet{Goodman90}. These observations show that there is a bimodal distribution
of polarization P.A., indicating that there are two large scale magnetic field
components along the line of sight.

NGC\,1333 IRAS\,4A (hereafter IRAS\,4A), 
a low-mass protostellar system, has become the textbook case
of a collapsing magnetized core: high angular submm polarimetric observations
have revealed that the magnetic field has an hourglass morphology at scales of
few hundred AUs \citep{Girart99, Girart06}.  This is the magnetic field
morphology predicted by theoretical models based on magnetically controlled
molecular core collapse \citep[e.g.,][]{Shu87,Mouschovias01}. Indeed, the
synthetic polarization maps constructed using models of collapsing magnetized
cores \citep{Galli93,Shu06} reproduced quite well the observations in IRAS~4A
\citep{Goncalves08}. In this context, it is worth mentioning that it is
still a question of ongoing debate whether magnetic fields or interstellar
turbulence plays a major role in the dynamical evolution of a molecular cloud
\citep[e. g.,][]{Crutcher09, Mouschovias10}.

In this paper, we report on one of the first scientific results obtained with  the
near-IR camera LIRIS \citep[Long-slit Intermediate Resolution Infrared 
Spectrograph:][]{Acosta03,Manchado04} in its polarimetric mode. The observations were
done using the $J$-band filter toward stars located relatively close to IRAS~4A
($\sim 4'$--$8'$). The fields were selected to avoid the most  active
star-forming portion of the NGC~1333 cloud, so the measured polarized light is
mainly due to dichroic absorption.
In order to ascertain the quality of the near-IR data, we also provide 
complementary $R$-band linear polarimetry obtained with the Observat\'orio do
Pico dos Dias toward the same region. 
The scientific goal of this work is to compare the magnetic field observed in
the IRAS\,4A molecular core with the larger scale field associated with the cloud
surrounding IRAS~4A. \citet{Girart06} have already done this comparison but using very
few distant stars ($\sim 14'$--$20'$) retrieved from the \citet{Goodman90}
survey.

 \section{Observations}

 \subsection{Near-infrared observations}\label{WHT}

The near-infrared observations were carried out in December 2006 and  December
2007 at the Observatorio del Roque de los Muchachos (La Palma,  Canary Islands,
Spain). The LIRIS camera, attached to the Cassegrain focus of  the 4.2 m
William Herschel Telescope, is equipped by a Hawaii detector  of $1024
\times 1024$ pixels optimized for the 0.8 to $2.5 \mu$m range.

\begin{figure}
\centering
\includegraphics[width=\columnwidth]{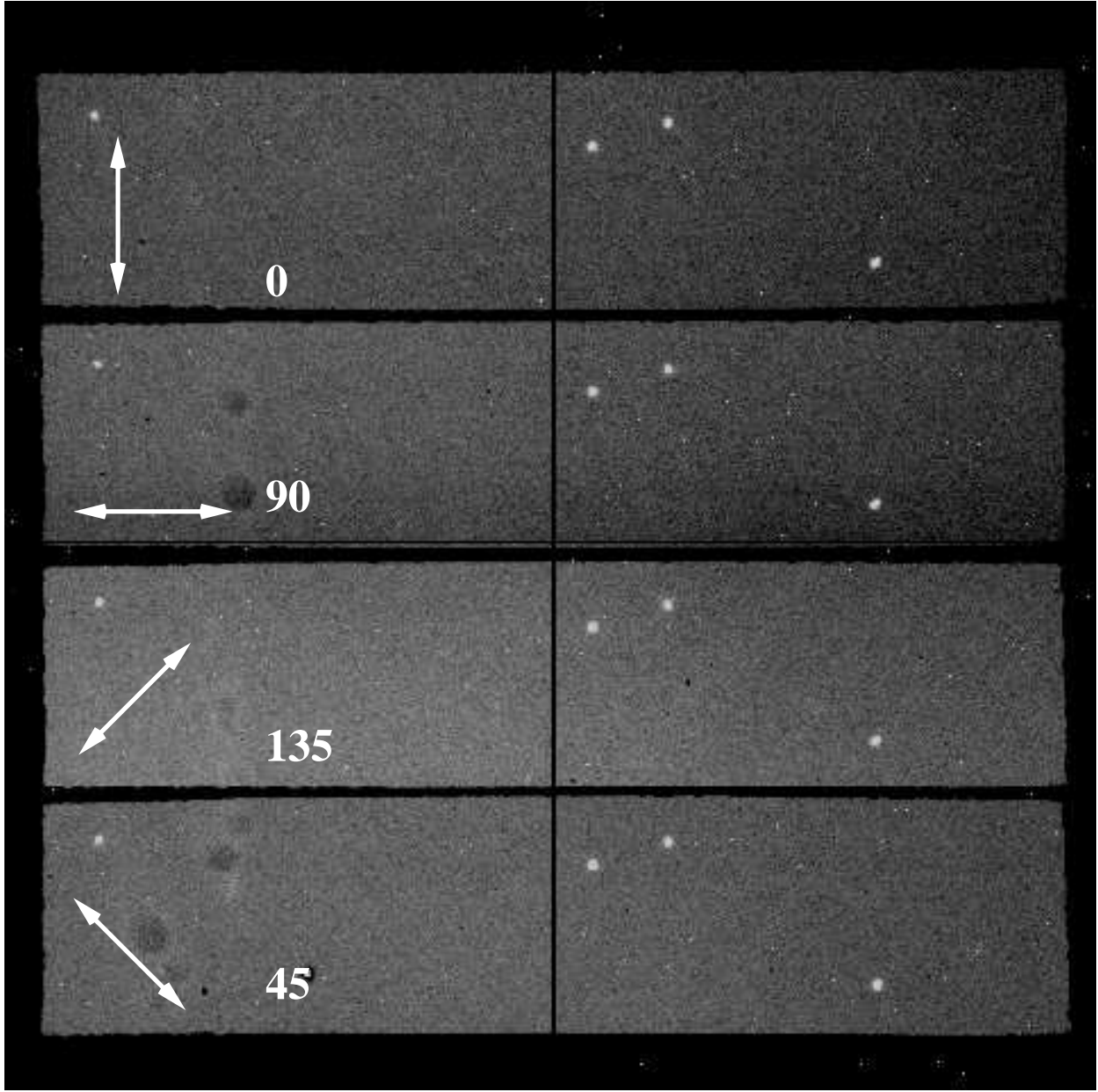}
\caption{A typical CCD image in polarization mode. The four strips correspond 
to the 0$\degr$, 90$\degr$, 135$\degr$ and 45$\degr$ polarization vectors 
from which the Stokes parameters are calculated.}
\label{imliris}
\end{figure}

LIRIS is capable of performing polarization observations by using a wedged double 
Wollaston device, WeDoWo, which is composed by a combination of two  Wollaston
prisms and two wedges \citep[see][for detailed description]{Oliva97}. In this
observing mode, 
the polarized flux is measured simultaneously at four different angles 
($0\degr$, $45\degr$, $90\degr$ and $135\degr$).
An aperture mask of $4' \times 1'$ is used in order to avoid  overlapping 
between the different polarization images. 
Figure~\ref{imliris} shows a typical LIRIS image in polarimetric mode.
The degree of linear polarization can thus be determined from data taken 
at the same time and with the same observing conditions.
In order to achieve accurate sky subtraction, a 5-point dither pattern 
was used. Offsets of about 20$^{\prime\prime}$ were adopted along the 
horizontal, long mask direction.  
During the 2006 and 2007 campaigns, we took seven and six exposures,
respectively, of 20~s per dither position.
The 5-point dither cycle was repeated  several times until completion of the
observation. The total observing time for  each field was 2800~s in 2006 and
2400~s in 2007.

\begin{figure}[t]
\centering
\includegraphics[width=\columnwidth]{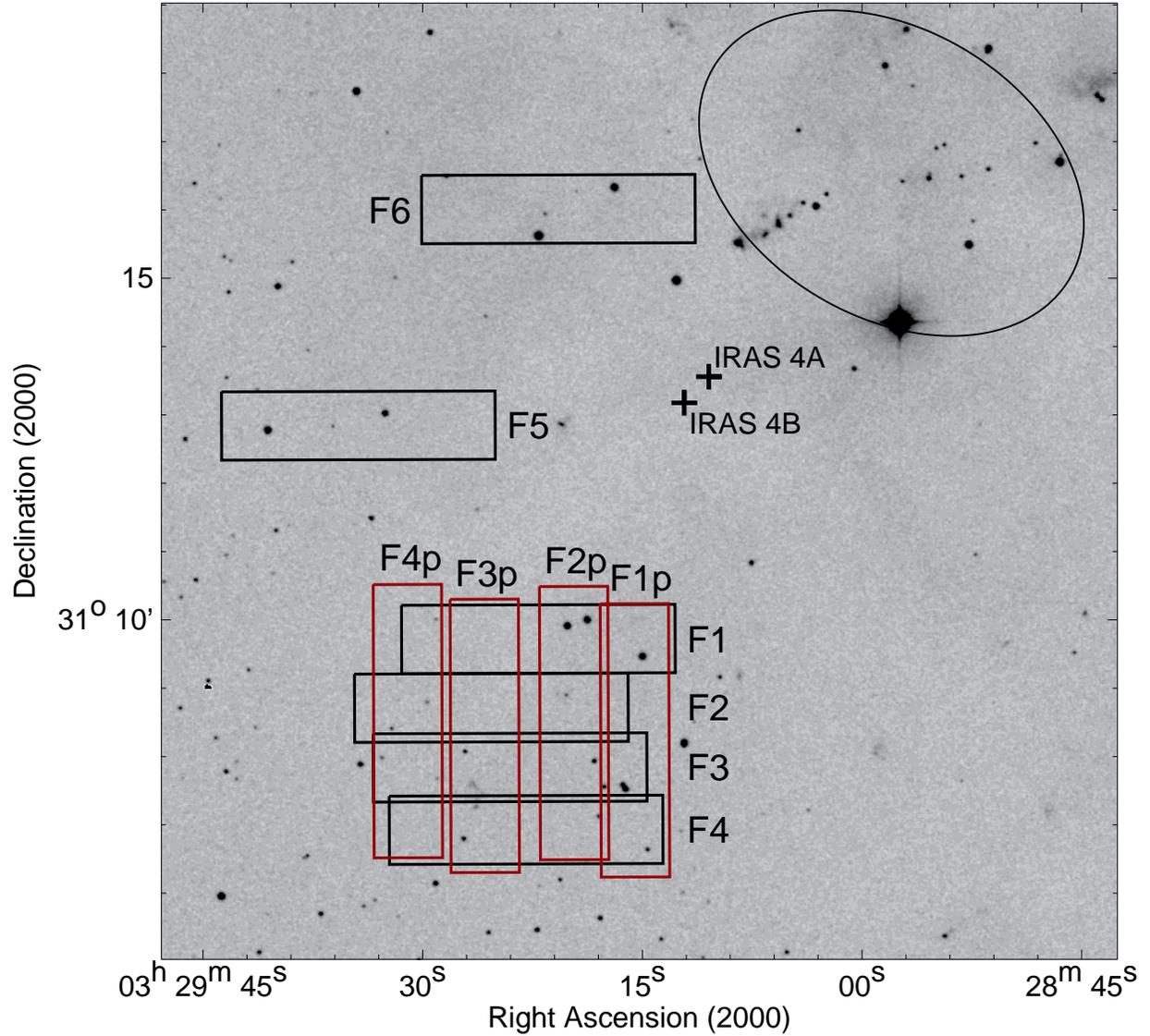}
\caption{DSS $R$-band image of our science targets. Black boxes are 
observed fields with rotator at 0$\degr$ while red boxes are observed fields 
at 90$\degr$. Crosses mark the positions of the protostars NGC 1333 IRAS 4A 
and NGC 1333 IRAS 4B. The ellipse in the upper right corner indicates the 
star-forming region, where no science targets were select in order to avoid 
polarization data due to dust scattering.}
\label{campos}
\end{figure}

We carried out $J$-band polarization observations of ten fields, six of them with 
the telescope rotator at 0$\degr$ and four  with the rotator at 90\degr\ 
(see Table\,\ref{targets}). Figure \ref{campos} indicates the observed fields
as  black and red rectangles, corresponding to observations with rotator at 
0\degr\ and 90\degr, respectively. 
We covered the  area surveyed by observing with
the rotator at 0\degr\ and 90\degr, except for the two upper fields. This procedure allows  us to compare both
data sets and, consequently,  to achieve higher precision in the  estimated
polarization parameters.   

\subsection{Optical observations}
\label{optobs}

The optical $R$-band linear polarimetry was performed using the 1.6 m telescope 
of the Observat\'orio do Pico dos Dias (LNA/MCT, Brazil) 
during observing runs
conducted  in 2007 and 2008. A specially adapted CCD camera composed by a
half-wave  rotating retarder followed by a calcite Savart plate and a filter
wheel was attached  to the focal plane of the telescope. The half-wave retarder
can be rotated in steps of 22\fdg5 and one polarization modulation cycle is
fully covered  after a complete 90$\degr$ rotation. The birefringence property
of the Savart  plate divides the incoming light beam into two perpendicularly
polarized  components: the ordinary and the extra-ordinary beams.  From the
difference in the measured flux for each beam one estimates the degree of
polarization and its orientation in the plane of the sky. For a
technical 
description of this polarimetric unit, we refer the interested reader to  the
work by \citet{Magalhaes96}. The obtained optical data is part of an  ongoing
large scale ($\sim 1$ square degree) survey whose results will be  discussed in
a forthcoming paper (Franco et al., in preparation). The area  covered by the
optical survey overlaps the portion of the sky observed in  near-IR, and in
order to make a comparative analysis of the results obtained at both
wavelengths, we included the optical results gathered for stars lying in the 
overlapped area in the discussion.

\section{LIRIS data reduction and calibration} 

The near-IR data reduction was performed using the  {\it lirisdr} package developed 
by the LIRIS team in the IRAF environment.\footnote{IRAF is distributed by 
the National Optical Astronomy Observatories, which are operated by the 
Association of Universities for Research in Astronomy, Inc., under cooperative 
agreement with the National Science Foundation.} Given the particular 
geometry of the frames (see Fig.\,\ref{imliris}),  
the first procedure was to slice the image into four frames.
Each set of frames corresponding to  a given polarization stage is processed
independently. The data reduction  process comprises sky subtraction,
flat-fielding, geometrical distortion  correction, and finally co-addition of
images after registering. A second  background subtraction was 
performed upon flat-fielded images in
order to avoid the residuals introduced  by the vertical gradient due to the
reset anomaly effect associated with the Hawaii arrays
\citep[e.g.,][]{Acosta06}.
An approximate astrometric solution was determined based on the image 
header parameters.  
 
\subsection{Photometry}

Aperture photometry of the field stars in each slice was obtained using the
task  Object Detection, available within Starlink Gaia software.\footnote{GAIA
is  a derivative of the Skycat catalogue and image display tool, developed as a
part  of the VLT project at ESO.} The aperture radius used was $\sim 4''$,
which  corresponds to 3 times the median seeing of the night. The background
was  extracted from an annulus with 
an inner radius of
$6^{\prime\prime}$ and an outer radius of $8^{\prime\prime}$. 
The astrometric solution of each  slice was tweaked using the
astrometric tools available within the Starlink Gaia  software. We used the
2MASS catalogue to perform the photometric and  astrometric calibrations. In our
sample, we reached $J$ magnitudes as faint  as $\sim$17. 
 As a final step, we identified the counterparts
of each object in the four slices in order to compute the polarization
properties.  In some cases, matching of stars observed with rotator at 0\degr\
and 90\degr\  was also necessary since some objects were present in both sets
of observations.

\subsection{Polarimetric analysis}
\label{pol_analysis}

Using the WeDoWo, we measured simultaneously four polarization states in each 
of the strips as,

\begin{eqnarray}
i_{0}[PA=0] &  =  & \frac{1}{2}\, t_{0}\, \mathrm{(I_* + Q_*)} \\
i_{90}[PA=0] & =  & \frac{1}{2}\, t_{90}\, \mathrm{(I_* - Q_*)}  \\
i_{45}[PA=0] & =  & \frac{1}{2}\, t_{45}\, \mathrm{(I_* + U_*)}  \\
i_{135}[PA=0] & =  & \frac{1}{2}\, t_{135}\, \mathrm{(I_* - U_*)} 
\end{eqnarray}
where $I_*$,  $Q_*$ and $U_*$ are the Stokes parameters of the object to be 
measured, and the factors $t_{[0,90,45,135]}$ represent the transmission for 
each polarization state. In this case, the normalized Stokes parameters can be 
determined by,

\begin{eqnarray}
\label{queq}
\mathrm{q_*} & = & \frac{i_{0}-i_{90}\, t_{0/90}}{i_{0}+i_{90}\, t_{0/90}}  \\
\mathrm{u_*} & = & \frac{i_{45}-i_{135}\, t_{45/135}}{i_{45}+i_{135}\, t_{45/135}} 
\end{eqnarray}
where the factors $t_{0/90}$ and $t_{45/135}$ measure the relative
transmission  of the ordinary and extraordinary rays for each Wollaston. 
These factors were  calibrated using non-polarized standards and
resulted in the values $t_{0/90} =0.997$  and $t_{45/135}=1.030$, with an
uncertainty of about 0.002 in both  cases.

The rotation of the whole instrument by 90\degr\ causes the exchange of 
the optical paths for the orthogonal polarization vectors. Now, the resulting 
polarization states are given by 

\begin{eqnarray}
i_{0}[PA=90] & = & \frac{1}{2}\, t_{0} (I_* - Q_*) \\
i_{90}[PA=90] & = & \frac{1}{2}\, t_{90} (I_* + Q_*) \\
i_{45}[PA=90] &  = & \frac{1}{2}\, t_{45} (I_* - U_*) \\
i_{135}[PA=90] & = & \frac{1}{2}\, t_{135} (I_* + U_*) 
\end{eqnarray}

This effect can be used in order to get a more accurate estimate of the Stokes 
parameters because the combination of both measurements, PA=0\degr\ and 90\degr, 
results in the cancelation of the transmission factors and reduces flat-field 
uncertainties. The normalized Stokes parameters are then computed by

\begin{eqnarray}
\label{qeqR}
\mathrm{q_*} &=& \frac{R_Q - 1}{R_Q + 1},\, \mathrm{being}\,\, R_Q^2 
	= \frac{i_{0}[PA=0]/i_{90}[PA=0]}{i_{0}[PA=90]/i_{90}[PA=90]} \\
\mathrm{u_*} &=& \frac{R_U - 1}{R_U + 1},\, \mathrm{being}\,\, R_U^2 
	= \frac{i_{45}[PA=0]/i_{135}[PA=0]}{i_{45}[PA=90]/i_{135}[PA=90]} 
\end{eqnarray}

Finally, after estimation of the $q$ and $u$ Stokes parameter, the degree of
linear  polarization and the position of polarization angle (measured eastwards
with  respect to the North Celestial Pole) are calculated as
\begin{eqnarray}
\label{pol}
p & = & \sqrt{q_*^{2} + u_*^{2}} 
\end{eqnarray}
\begin{eqnarray*}
\label{theta}
\theta & = & \frac{1}{2}~ \mathrm{tan}^{-1} \left(\frac{u_*}{q_*}\right).
\end{eqnarray*}

Flux errors in $i_{0}$, $i_{90}$, $i_{45}$ and $i_{135}$  are dominated by
photon  shot noise while the theoretical error in polarization fraction was
estimated  performing error propagation through the previous equations. In
addition,  we calculated the errors in $p$ using a Monte Carlo method, which
returned  values 
similar to those estimated from error propagation.
The 1$\sigma$ uncertainty in $\theta$ was
estimated ({\it i}) by  applying the relation derived by \citet{Serkowski74}
using standard error propagation, 
that is, $\sigma_{\theta} =  28\degr.65~\sigma_{p}/p$, when $p/\sigma_p
\ge 5$; or ({\it ii}) graphically  with the aid of the curve proposed by
\citet{NKC93} when $p/\sigma_p < 5$.
    
\begin{figure}[t]
\centering
\includegraphics[width=\columnwidth]{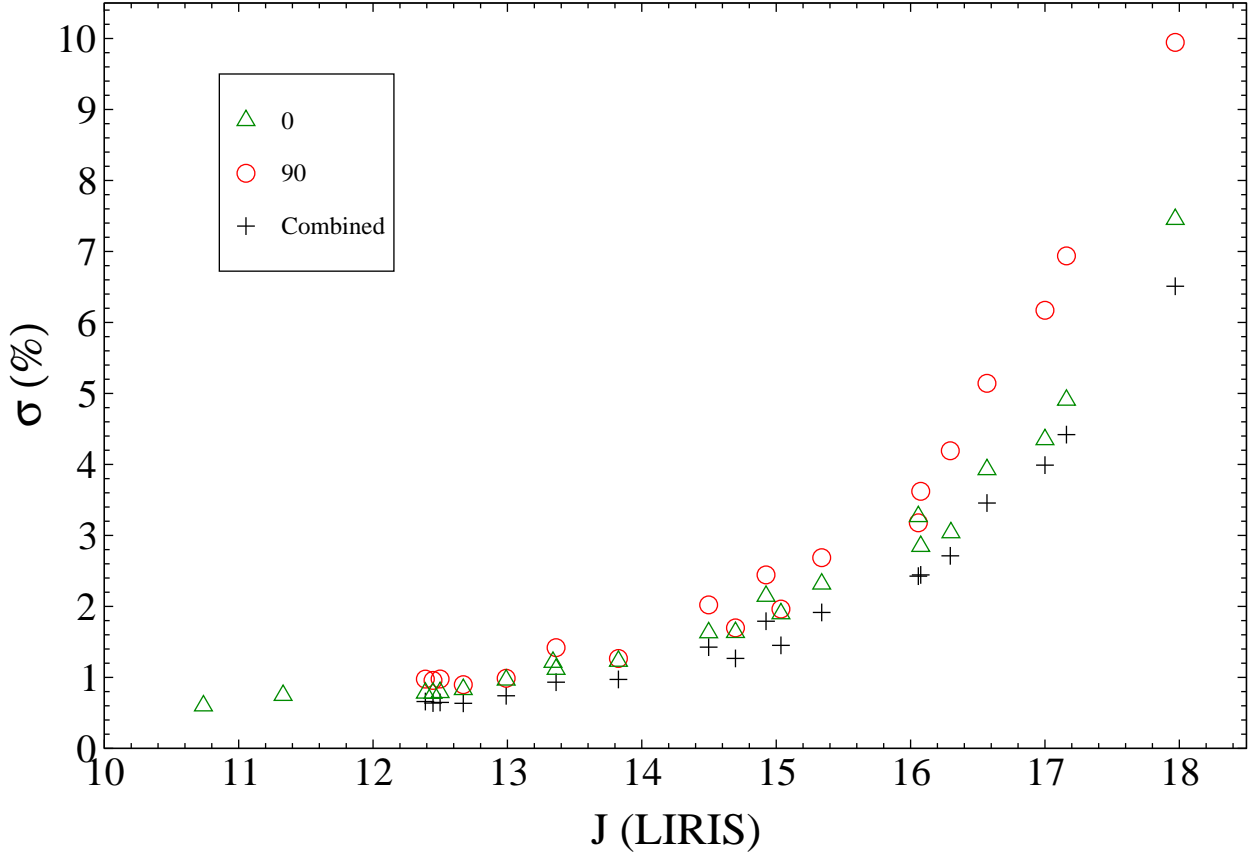}
\caption{Distribution of the polarization uncertainty ($\sigma_{P}$) with 
respect to the $J$ magnitude obtained for each field star with the telescope rotator 
at 0$\degr$ (green triangles), 90$\degr$ (red circles) and a combination 
of both (crosses). 
This plot contains only stars whose signal-to-noise ratio of the combined setup 
(crosses) is better than 1 (except for three stars that also have a 
signal-to-noise ratio better than 1 but which were only observed with the 
rotator at 0\degr, see Table \ref{tab:IRdata}).
Note that uncertainties are lower when the combination of images taken 
with the telescope rotator at 0$\degr$ and 90$\degr$ is used. The large 
discrepancy observed between 0$\degr$ and 90$\degr$ errors for some stars 
are due to the distinct observation epochs of each data set.}
         \label{shot}
\end{figure}

Figure~\ref{shot} shows 
the polarization uncertainty as a function of 
the $J$-band magnitude achieved with our LIRIS observations.
The observed distribution
suggests that the uncertainties are dominated by photon shot noise, as
expected  for a sample collected with fixed exposure time.  
As expected, the
uncertainties decrease when the data taken at 0\degr\ and 90\degr\ are
combined. There is a natural limit which is due to the uncertainty bias
when measuring low levels of polarization. Bias in the 
degree of linear polarization ($p$) comes from the fact that this quantity is 
defined as a quadratic sum of $q$ and  $u$, which produces a non-zero
polarization estimate due to the uncertainties in their measurement \citep[for
a detailed discussion see for
instance,][]{Simmons85,Wardle74}. In order to remove the polarization bias 
and compute the true polarization, we used the
prescription  proposed by \citet{Simmons85} for low polarization stars. The 
true polarization degree can be approximated by the expressions $p_{true} = 0$ 
if $p_{obs}/\sigma_{p} < K_a$, otherwise $p_{true} = (p_{obs}^2 - \sigma_{p}^2 
\cdot K_a^2)^{1/2}$. We adopted $K_a=1$, which corresponds to the estimator 
defined by \citet{Wardle74}.  

\subsection{Standard stars}

Observations of polarized and unpolarized standard stars were taken in order
to  calibrate the instrumental characteristics of LIRIS in its polarimetric
mode. Table~\ref{sstars} summarizes the general information
for these stars: columns 1 to 8 indicate their name, equatorial coordinates,
type, polarization degree and position angle, the filter used for the
polarization measurements and the reference, respectively.
Unpolarized standard stars were observed to check for any possible
instrumental  polarization and for systematic errors in our polarimetry. The
unpolarized stars  G191B2B and BD+28d4211 were observed with rotator at 0\degr\
and 90\degr. 
The polarized intensity measured for the two unpolarized standards was very
small (see Table \ref{unpol}): The measured normalized Stokes $q$ and $u$ were
0.051\% and 0.226\%, respectively, with the rotator at 0\degr, and $-0.117$\%
and  0.119\%, respectively, with the rotator at 90\degr. Table~\ref{unpol}
shows the observed polarization degree before and after bias correction for the
two unpolarized standards. The measurements taken in the two epochs for 
BD+28d4211 give consistent values. 
We applied the method proposed by \citet{Simmons85} for  a 99\% confidence
level for the observed unpolarized standards. This resulted  in a small, if
 any, instrumental polarization. 
 
The polarized standard star CMa R1 No. 24 was observed in order to verify  the
zero point of the polarization position angles. Table \ref{tab:pol} summarizes 
the results obtained for the four measurements conducted for this object. As 
expected, high quality data are less sensitive to biasing, and the unbiased 
polarization has basically the same values of the observed polarization. 
Taking into account the uncertainties, we see that our $J$-band data  match the
result obtained by \citet{Whittet92}. The difference between the
average P.A. obtained for these four measurements  and that obtained by
\citet{Whittet92} is $\sim 6\degr$, which is very close to 
the statistical deviation of our measurements thus discarding any further 
correction for the zero angle calibration.

\section{Polarization properties}

\subsection{Infrared data}
\label{sec_irdata}

\begin{figure}[t]
   \centering
  \includegraphics[width=\columnwidth]{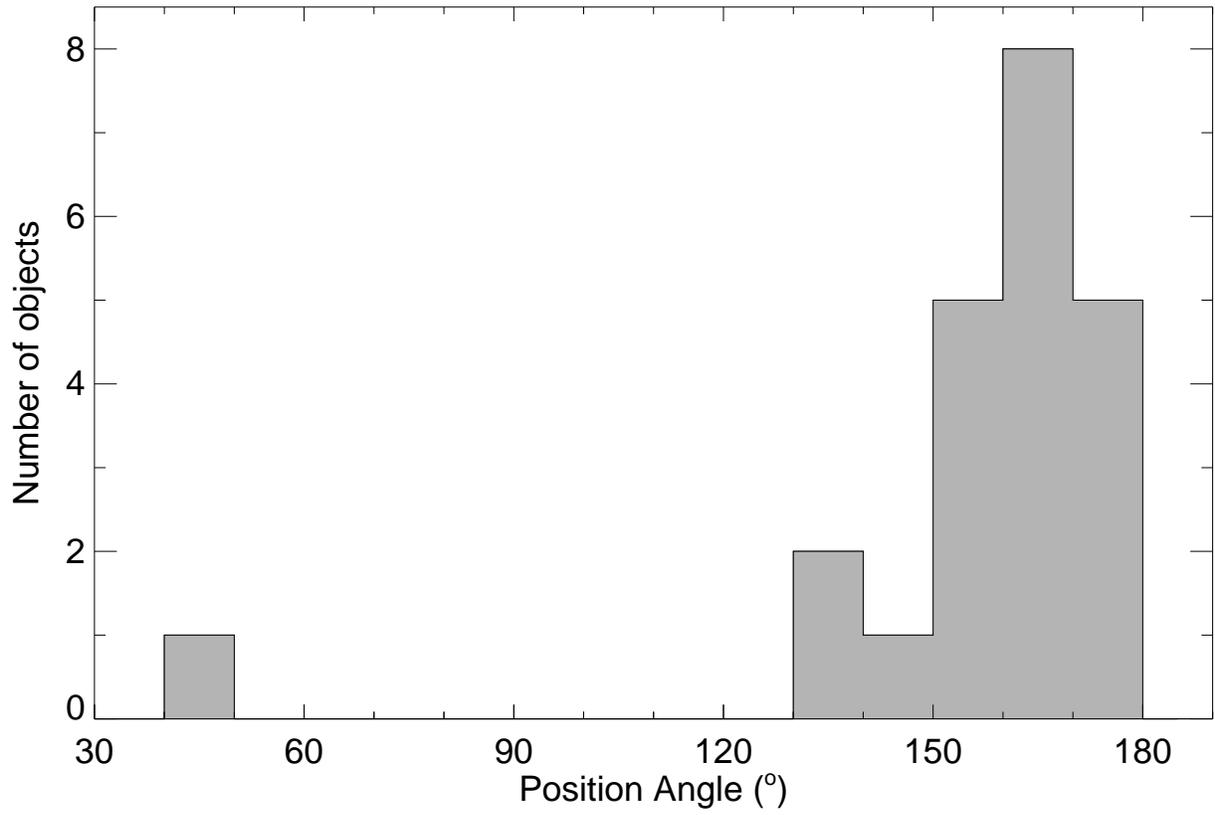}
     \caption{Distribution of polarization angles of the near-IR data. The  histogram is binned
     in 10$\degr$.}
         \label{histPA}
   \end{figure}

Table \ref{tab:IRdata} contains a summary of our near-IR 
and optical polarization data for stars 
with a signal-to-noise in the polarization 
intensity higher than unity. Column 1
gives the star's identification number in our catalogue. Columns 2 and 3 show
the equatorial coordinates. Column 4 gives the $J$-band magnitude. Columns 5 to
12 show the polarization degree and the polarization P.A. (with their
uncertainties) for the $R$ and $J$ bands. The  last two columns indicate the
rotator position used to acquire the near-IR data, and the object type, respectively.
Figure~\ref{histPA} shows that, excluding star 13, the polarization P.A.
distribution measured in the $J$-band is quite narrow with a mean position angle
of 160\degr\ and a standard deviation of only 12$\degr$. We note that the $J$-band
polarization uncertainties may be overestimated. First,  the aforementioned
standard deviation is about half the mean 1-$\sigma$ uncertainty of the
polarization P.A., and second, there is a very good agreement between the
near-IR and optical data (see next section).

\begin{figure}
   \centering
   \includegraphics[width=.8\columnwidth]{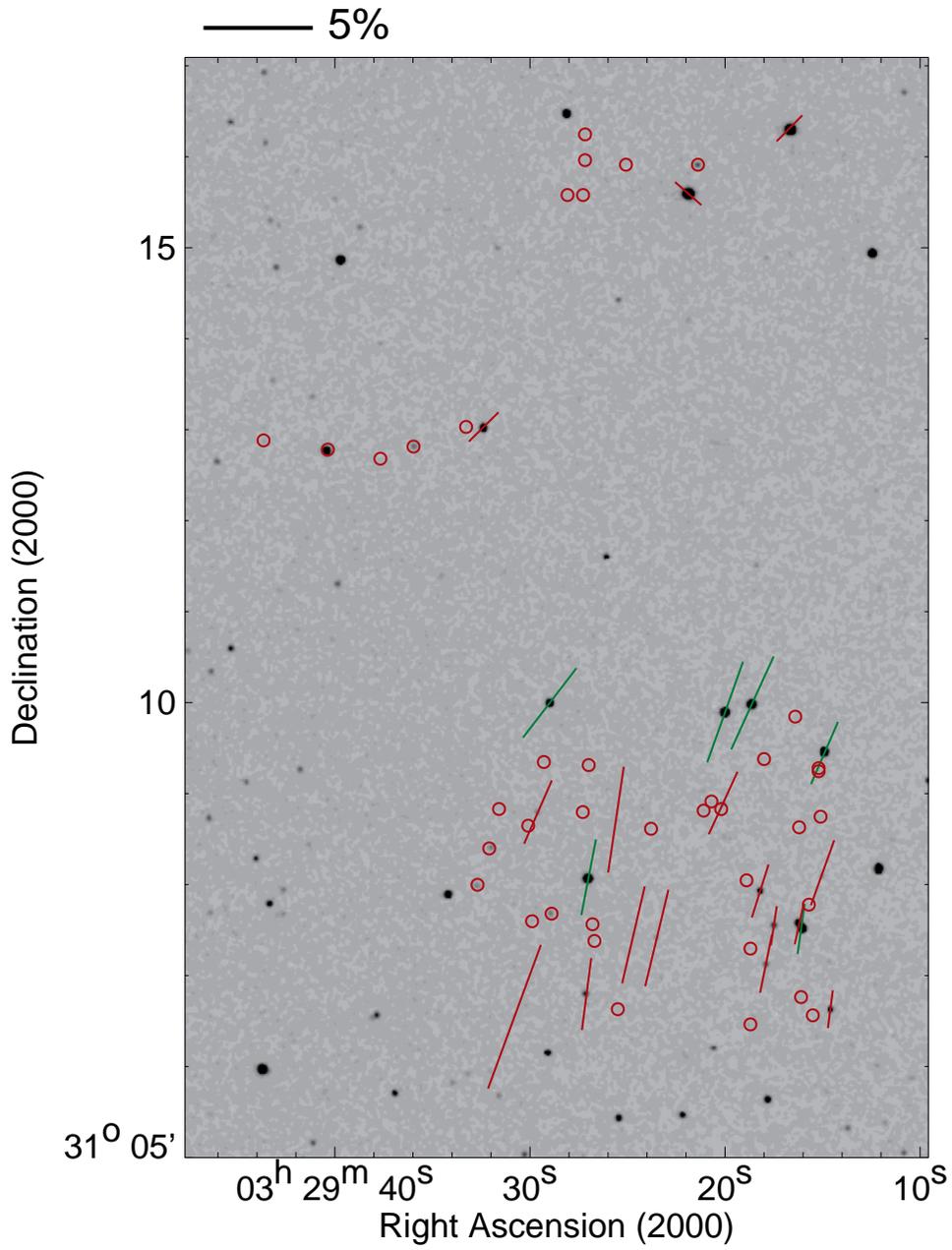}
      \caption{$J$-band polarization vectors in NGC 1333 plotted over a 2MASS 
      $J$-band image. Vector length scale is shown on the upper left
      corner. Green vectors indicate stars with P/$\sigma_{\mathrm{P}} > 3$ 
      while red vectors have $1 < \mathrm{P}/\sigma_{\mathrm{P}} < 3$. Open 
      circles indicate positions of observed objects with P/$\sigma_{\mathrm{P}} < 1$.
Some of the detected polarized stars in the $J$-band does not have a 
2MASS counterpart, indicating that the obtained LIRIS data probe deeper visual extinctions in the cloud than the 
2MASS.}
         \label{IRpolmap}
   \end{figure}

Figure\,\ref{IRpolmap} shows the spatial distribution of the near-IR
polarization vectors overlaid on the 2MASS $J$-band image. 
The polarized stars with a declination below $31\arcdeg
10'$ have larger  polarization degrees than those above this value.
This subsample comprises most of stars with mean P.A. 
$\simeq$ 160$\degr$. Star number 13 is the only object in our catalogue
presenting intrinsic polarization  
(see section \ref{pol_YSOfor}).

\subsection{Comparison with optical data}
\label{comp_opt}

\begin{figure}[t]
\centering
\includegraphics[width=\columnwidth]{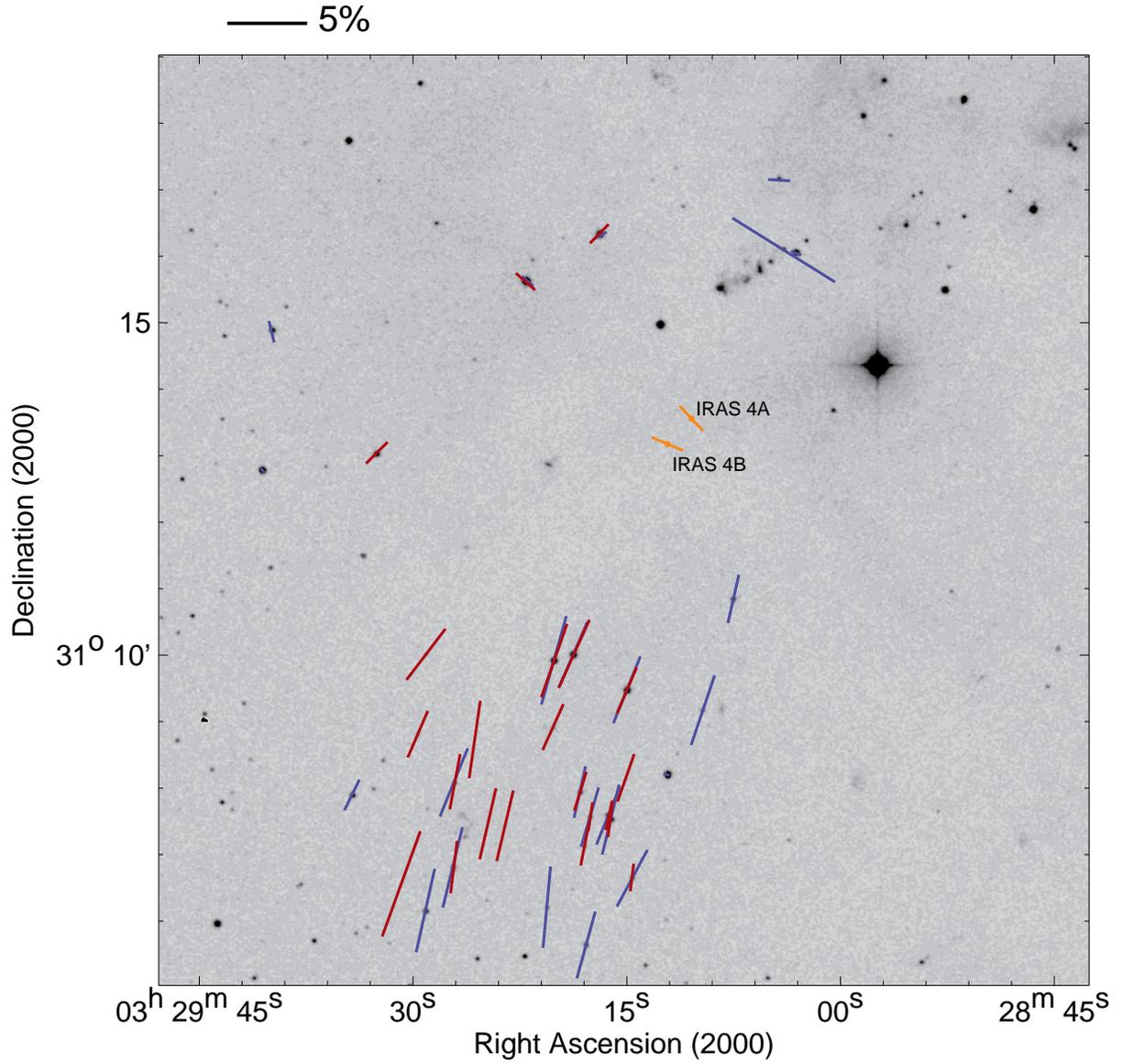}
\caption{Comparison between optical (blue vectors) and near-IR (red vectors) data. 
The polarimetric map is plotted over a DSS $R$-band image. The vector length 
scale is shown on the upper left corner. Orange vectors represent the averaged 
magnetic field of IRAS\,4A and IRAS\,4B, as obtained by submillimeter observations 
of \citet{Attard09}.}
\label{mapaVIS_IR}
\end{figure}

\begin{figure}
\centering
\includegraphics[width=\columnwidth]{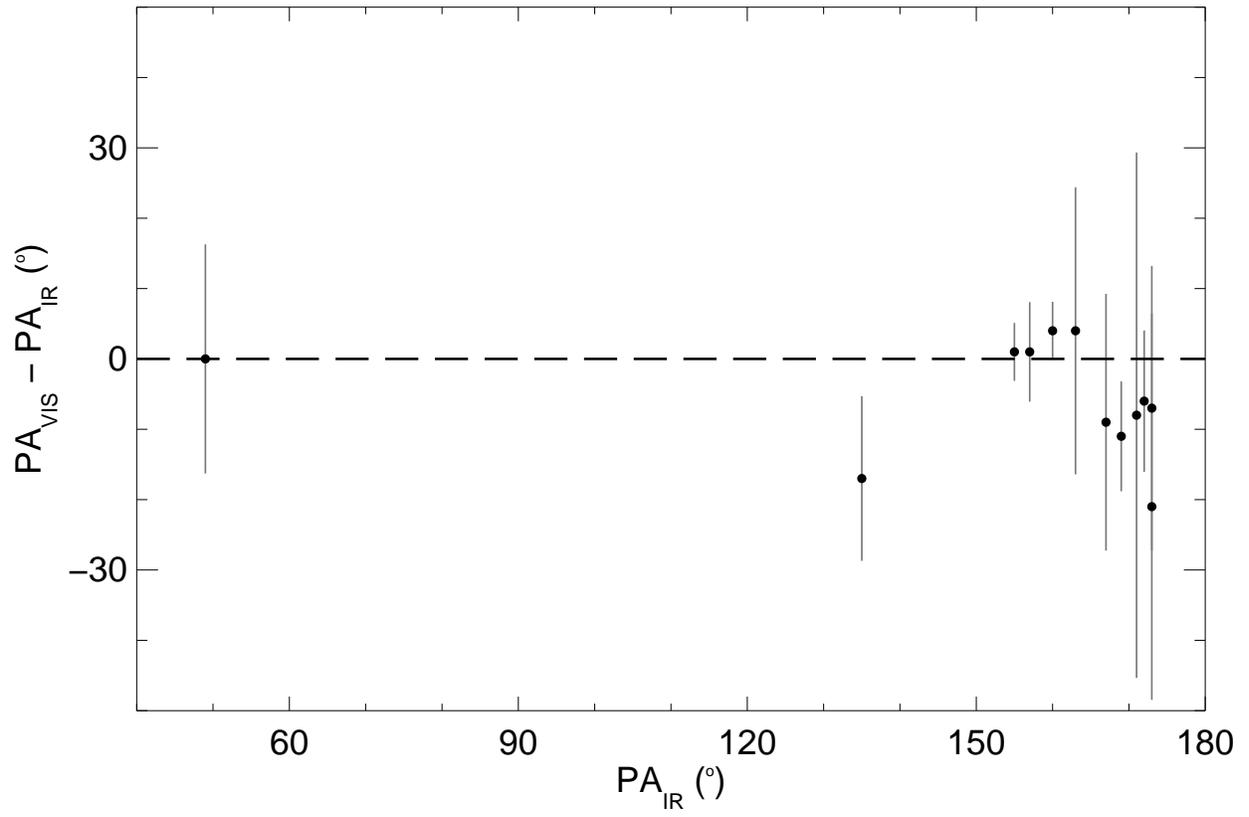}
\caption{Comparative diagram of the position angles obtained for the optical ($R$-band) and near-IR ($J$-band) data sets.}
\label{PAcomp}
\end{figure}

Previous
optical polarimetric observations performed toward the field of view shown in
Fig.~\ref{IRpolmap} detected only two polarized stars of the $J$-band sample,
stars number 2 and 13  \citep{Vrba76, Menard92}, and our data are in good agreement
with them.
The $R$-band polarimetric sample 
has 12 stars in common with our near-IR data. Figure \ref{mapaVIS_IR} shows the
polarization vectors in both bands plotted over a DSS image.
There is noticeably good agreement between the two polarization data sets (see
also Fig.~\ref{PAcomp}). Thus, the mean value of the P.A. difference between
the $R$ and $J$-band for the 12 stars is 6\fdg5.

\begin{figure}
\centering
\includegraphics[width=\columnwidth]{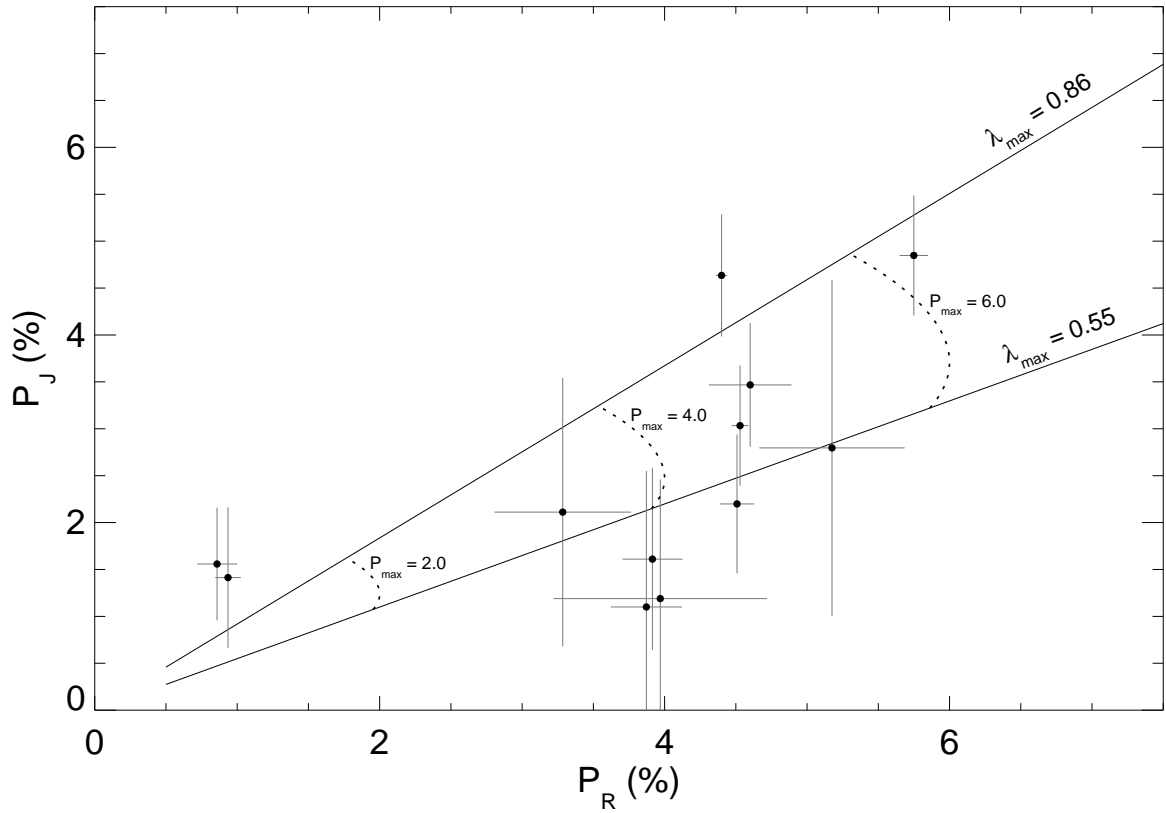}
\caption{Spectral Energy Distribution of the observed linear polarization in 
near-IR and visible. Solid lines indicate constant $\lambda_{max}$ of 0.55 and 
0.86 $\mu$m from bottom to top, respectively. Dashed lines represent constant 
$p_{max}$ of 2, 4 and 6\% from the origin going outwards, respectively.}
\label{sed_plot}
\end{figure}

Figure\,\ref{sed_plot} represents the resulting P$_{NIR}$ vs.
P$_{visible}$ diagram for the 12 stars with both $J$ and $R$ polarimetric
measurements.  The wavelength at which the polarization is
highest, $\lambda_{max}$, is related to the mean size of the interstellar grains
responsible for producing the observed polarized light \citep{Serkowski75,
McMillan78}.
The typical value of $\lambda_{max}$ observed for the diffuse interstellar
medium is 0.55~$\mu$m \citep{Serkowski75}. 
However, in molecular clouds  $\lambda_{max}$ appears to change with the visual
extinction \citep{Andersson07}. In the case of NGC 1333, it has been shown that
$\lambda_{max}$ changes between 0.66 and 0.89~$\mu$m \citep{TTC80,Cernis90, Whittet92}.
The two solid lines in Fig.~\ref{sed_plot} show the expected relation for
$\lambda_{max} = 0.55$ and 0.86~$\mu$m. 
The data do not show any preferential regime of maximum polarization 
due to the low statistics. Polarization observations over a wider range of 
wavelengths are necessary to refine this characterization.

\section{Polarization from YSOs and foreground stars}
\label{pol_YSOfor}

Our near-IR and optical maps are characterized by a uniform component
predominant to the south of the IRAS 4A/4B double system 
(see Fig.~\ref{mapaVIS_IR}).  However, north of IRAS 4A/4B the few detected 
polarized stars have a broader angle distribution. Most of these stars are
likely  young stellar objects (YSOs). In these cases, the polarization is
produced by  intrinsic scattering within circumstellar disks rather than by
interstellar absorption.
Several authors have studied the physical properties of YSOs by means of 
polarimetry \citep[e.g.,][]{Brown77,Mundt83} and, in general, observations 
suggest that near-IR polarization vectors, when produced by single  scattering,
are oriented perpendicularly to optically thin disks, while multiple  scattering
within optically thick disks generates polarization vectors whose P.A. are
parallel to their long axis \citep{Angel69,Bastien90,Pereyra09}. 

The YSOs possibly showing intrinsic polarization in the near-IR and/or
$R$-band are listed below:

\begin{itemize}
\item LkH$\alpha$ 271 
(star n. 13 of Table \ref{tab:IRdata}) is a
Classical T Tauri star \citep{Lada74}. The near-IR polarization angle and
degree are in excellent agreement with the $R$-band data. 
Previous observations showed that the polarization varies considerably, which 
has being interpreted as arising from outbursts or inhomogeneities in a 
circumstellar shell within an optically thick circumstellar disk
\citep{Tamura88, Menard92}.

\item 
SVS 13A (star n. 24) was observed only in the $R$-band. The obtained polarization
is in excellent agreement with the value previously measured in the $K$-band  
\citep[$P_{K}$ = 7.2 $\pm$ 0.9\% and P.A. = 56 $\pm$ 4$\degr$,][]{Tamura88}.
This object is a well-studied source \citep{Rodriguez02,Anglada04,Chen09} which
powers a bipolar and collimated  outflow associated with the well-known
Herbig-Haro objects HH 7-11  \citep{Herbig74,Strom74}. 
The orientation of the outflow is roughly perpendicular to the polarization P.A., which
suggests that the disk is optically thick.

\item 
2MASS J03290289+3116010 (star n. 23) is close to SVS\,13A and has a very  low
degree of polarization. According to the SIMBAD Astronomical Database,  this is
a bright  ($J \simeq 12.8$)  K-type star and it is possibly a foreground star.
In fact, previous spectral analysis and photometric studies place  this star at
a distance of only $\sim$50 pc from the Sun \citep{Aspin94,Aspin03}.
 
\item 
ASR 8 (star n. 25) is classified as a brown dwarf by SIMBAD. However,
an extensive  survey on the evolutionary state of stars in NGC 1333 identifies
this object as a T Tauri star with a mass of 0.7\,M$_{\odot}$ \citep{Aspin03},
which is reinforced by the  presence of X-ray  emission \citep{Getman02}. We
therefore attribute the optical polarization measured for this star due to
intrinsic scattering. 

\end{itemize}

In addition, there are two 
bright infrared stars ($J \lesssim 13.0$) with low polarization (stars n.~28
and  34 in Table\,\ref{tab:IRdata}) that are
apparently not associated with YSOs, as no star formation or nebulosity 
signs has been reported in the literature. Their 2MASS color  indices suggest
that they may be unreddened M-type dwarf stars, which is also corroborated
by the low degree of polarization. We therefore consider these objects
to be foreground stars.

\section{The magnetic field in NGC 1333}

\subsection{The distribution of dust and molecular gas in NGC\,1333}
\label{gasdust}

The most detailed picture of the distribution of gas and dust in the Perseus 
cloud has been provided by the COMPLETE project \citep{Ridge06a, 
Pineda08}, a survey of near/far-infrared extinction data, and of atomic,
molecular,  and thermal dust continuum emission obtained over a
large area.  These data show a wide range of visual  magnitudes for NGC 1333,
and a non-Gaussian CO spectral profile consistent with  multi-velocity
components.  These results are consistent and likely  related to a layered cloud
structure along the line of sight, which was first  proposed by 
\citet{Unge87}.  Interstellar extinction studies of field stars toward NGC 1333
also  suggest at least two components in the line of sight at different
distances  toward NGC~1333 \citep{Cernis90}. 

According to column density maps of the Perseus cloud \citep{Ridge06b}, the
region studied here lies in the lower density envelope of NGC 1333.   Maps of
high density molecular tracers (N$_{2}$H$^{+}$, HCO$^{+}$) as well as of  the 
870~$\mu$m dust emission, show that around IRAS~4A the dense gas has a 
filamentary distribution oriented in the NW--SE direction, with the long axis
positioned at $\simeq$142\degr \citep{Sandell01, Olmi05, Walsh07}.

\subsection{The field morphology as traced by the diffuse gas}

The near-IR and optical polarization vectors of the background stars shown in 
Fig.~\ref{mapaVIS_IR} trace the POS component of the magnetic field associated 
with the lower density envelope around IRAS~4A/4B. South of these sources, where we 
have most of the polarization sample, the magnetic field has a direction of 
$\simeq 160\degr$. The observed configuration is consistent with the results 
obtained at much larger scale by \citet{Goodman90} and \citet{Tamura88}. 
According to the COMPLETE survey \citep{Ridge06a} the polarization was measured 
toward regions with a visual extinction of 4 to 5 mag.

\begin{figure}
\centering
\includegraphics[width=\columnwidth]{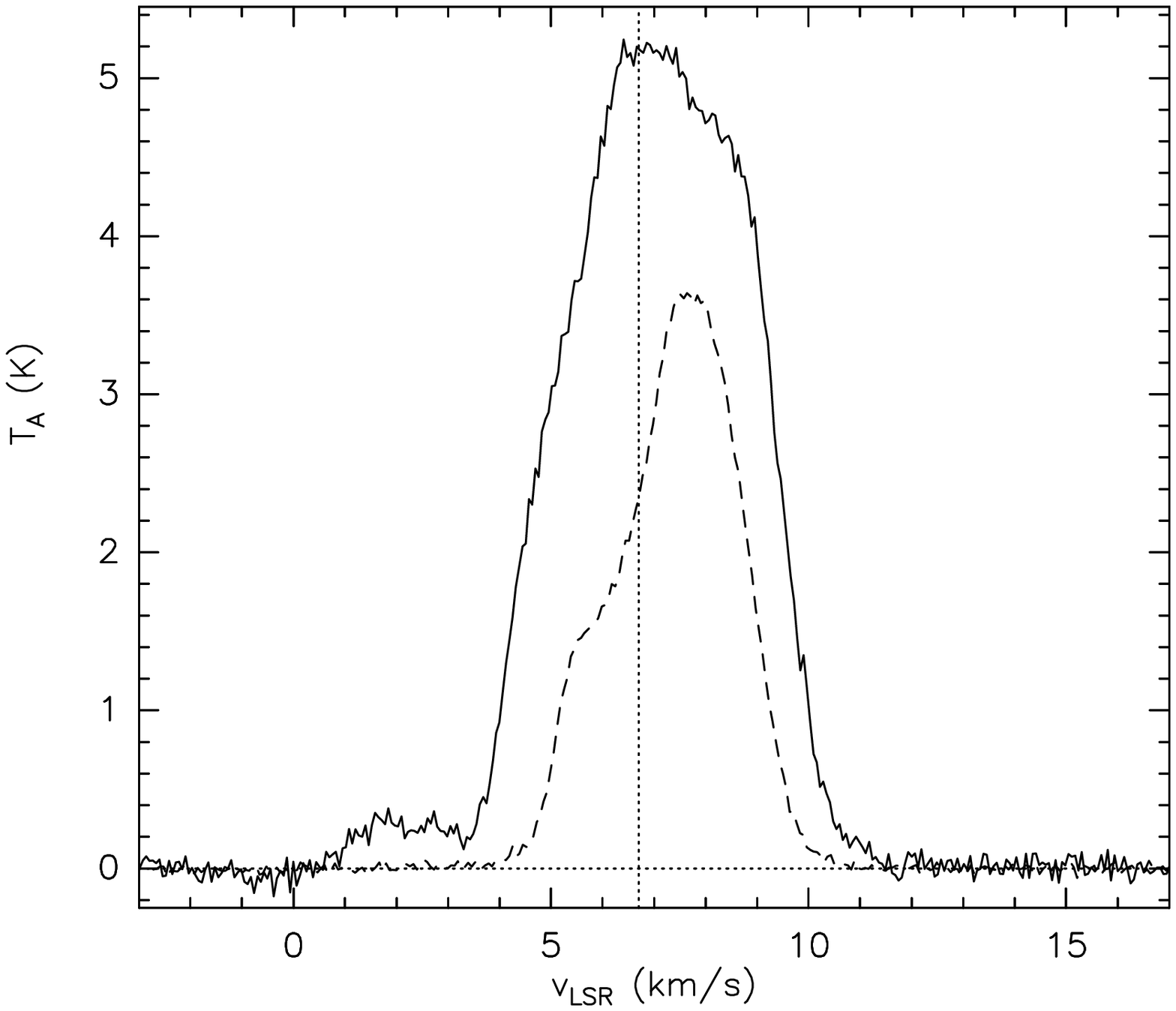}
\caption{Averaged spectra of the $^{12}$CO 1--0  (solid line) and the $^{13}$CO 1--0
(dashed line) lines obtained over a region of about $5\arcmin$ centered at the 
position $\alpha$(J2000)=$3^{\rm h} 29^{\rm m} 24^{\rm s}$and 
$\delta$(J2000)=$31\degr 8'$. This region covers the F1--F4 and F1p--F4p observed
fields with the WHT (see \S~\ref{WHT}). The CO spectra was
retrieved from the COMPLETE data archive \citep{Ridge06a, Pineda08}. 
The dotted vertical line shows the systemic velocity of the IRAS 4A 
core \citep{Choi01}.}
\label{COcomplete}
\end{figure}

The magnetic field orientation derived from our data is roughly parallel to the 
dense filamentary structure associated with IRAS~4A \citep{Walsh07,Sandell01}. 
However, the submm polarization maps towards IRAS~4A and IRAS~4B show that the
magnetic field within the filament is approximately perpendicular to the 
filament's major axis \citep{Girart99, Girart06, Attard09}, and is therefore 
perpendicular to the magnetic field direction traced by our optical and 
near-IR data.  
The single-dish submm polarization map from \citet{Attard09} around IRAS~4A 
is associated with visual extinctions as low as $\sim$10 magnitudes, which is
a typical value for near-IR extinction data.
Therefore, the submm and near-IR/optical data seem to reveal substantial 
changes in the magnetic field topology between the dense filament and the 
diffuse molecular envelope that surrounds it. Such a sharp twist in the field 
is hard to explain by means of structural changes in the magnetic field 
only, because within the observed field, the position angle of the optical 
and near-IR polarimetric data is quite uniform (see Fig.\,\ref{histPA}). 
Instead, the two data sets may be simply tracing distinct gas components. 
As explained in \S~\ref{gasdust}, there is observational evidence of a 
multi-component structure for the NGC 1333 molecular cloud. 
Figure~\ref{COcomplete} shows the $^{12}$CO and $^{13}$CO spectra extracted 
from a box containing the region studied here. These spectra show at least 
three distinguishable velocity components: a faint 
emission centered at  $v_{\rm LSR} \simeq 2$~\kms\ (seen more clearly in the  
$^{12}$CO data), the peak of the $^{13}$CO data centered at $\sim$7.6~\kms\ and 
the peak of the $^{12}$CO data at $\sim$6.7~\kms.  This last component 
has the same $v_{\rm LSR}$ of the IRAS 4A dense core \citep{Choi01}. 
Therefore, whereas the submm polarization measurements trace only the
molecular cloud component associated with the IRAS 4A dense core, 
the near-IR and optical polarimetric data are probably tracing the mean 
magnetic field of the different velocity molecular cloud components observed 
in the CO maps. Nevertheless, further observations are needed in order to 
obtain a more complete description of the magnetic field in this region. 
           
\section{Conclusions}

We have carried out one of the first polarimetric observations in the $J$-band 
collected with the WHT/LIRIS infrared camera.
We also present $R$-band linear polarimetry obtained at the Observat\'orio do 
Pico dos Dias.  We observed an area of $\sim$\,6$^{\prime} \times 4^{\prime}$ 
around the NGC\,1333 IRAS~4A/4B protostellar system. 
The main conclusions of this work are:

\begin{itemize}
\item  The infrared polarization map derived for the surveyed area is 
highly consistent with the optical map obtained with a 
different telescope and observational technique.
Therefore, the near-IR polarimetric capabilities of LIRIS have proved to be 
scientifically trustworthy for the astronomical community, and 
assure this 
mode will be useful for gathering measurements of objects experiencing high
interstellar extinction inaccessible to optical instruments.

\item The polarization map obtained for the surveyed area is dominated
by a well-ordered component produced by dichroic interstellar absorption. 
However, there are objects, some of them catalogued as YSOs, that show a 
transversal component which may be generated by internal scattering 
within circumstellar disks.
  
\item
The magnetic field 
morphology traced by the near-IR/optical map is almost perpendicular 
with respect to the field morphology obtained with the submillimeter data 
toward the dense molecular core around IRAS\,4A/4B. The near-IR/optical 
polarimetric data trace the field morphology of the diffuse molecular gas, 
which is known to have a multi-velocity structure. That is, the 
observed resulting magnetic field direction is probably the averaged 
magnetic field over several distinct velocity components of the cloud. 
CO molecular data obtained for this line of sight show non-Gaussian 
line profiles that are consistent with this hypothesis.
\end{itemize}

\begin{acknowledgements}
FOA acknowledges the hospitality of the Instituto de Astrof\'{\i}sica de 
Canarias, where part of this work was developed. 
The authors thank the staffs of
the Observatorio del Roque de los Muchachos and Observat\'orio do Pico dos
Dias for their hospitality and invaluable help during the observing runs. 
We also appreciate Terry Mahoney's help with the manuscript.
We made extensive use of NASA's Astrophysics Data System (NASA/ADS) and 
the SIMBAD database, operated at CDS, Strasbourg, France. CO spectra were 
retrieved from the COMPLETE Survey of Star-forming Regions 
\citep{Goodman04,Ridge06a}. This publication makes use of data products 
from the Two Micron All Sky Survey, which is a joint project of the University 
of Massachusetts and the Infrared Processing and Analysis Center/California 
Institute of Technology, funded by the National Aeronautics and Space 
Administration and the National Science Foundation. GAPF acknowledges a 
grant from Fundaci\'on Carolina (Spain). This research has been partially 
supported by AYA2008-06189-C03 and AYA2004-03136 (Ministerio de Ciencia 
e Innovaci\'on, Spain), CEX APQ-1130-5.01/07 (FAPEMIG, Brazil), 
CNPq (Minist\'erio da Ci\^encia e Tecnologia, Brazil) and 2009SGR1172 
(AGAUR, Generalitat de Catalunya).
\end{acknowledgements}

\break

\begin{deluxetable}{c c c c c}
\tablecaption{Log of the observations\tablenotemark{a}\label{targets}}
\tablewidth{0pt}
\tablehead{Target & $\alpha_{2000}$  & $\delta_{2000}$  & Obs. date & Rotator  \\ 
ID & (hh:mm:ss.ss) & (dd:mm:ss.ss) & &  ($\degr$)}
\startdata                   
   F1 & 03:29:22.01 & +31:09:42.12  &   2006 Dec 26 & 0 \nl
   F2 & 03:29:25.24 & +31:08:41.88  &   2006 Dec 26 & 0 \nl
   F3 & 03:29:23.98 & +31:07:49.41 &   2006 Dec 26 & 0 \nl
   F4 & 03:29:22.90 & +31:06:54.62  &   2006 Dec 26 & 0  \nl
   F5 & 03:29:34.26 & +31:12:50.08  &   2007 Dec 13 &  0 \nl
   F6 & 03:29:20.52 & +31:15:59.63  &   2007 Dec 13 & 0 \nl
   F1p & 03:29:15.44 & +31:08:12.82 &   2006 Dec 26 & 90 \nl
   F2p & 03:29:19.59 & +31:08:28.20 &   2007 Dec 13 & 90 \nl
   F3p & 03:29:25.70 & +31:08:17.24 &   2007 Dec 13 & 90 \nl
   F4p & 03:29:30.95 & +31:08:30.44 &   2007 Dec 13 & 90
\enddata 
\tablenotetext{a}{The night of 2006 December 27 had very limited 
weather conditions and only calibrators were observed.}              
\end{deluxetable}

\begin{deluxetable}{c c c c c c c c} 
\tablewidth{0pt}
\tabletypesize{\scriptsize}
\tablecaption{Standard stars\label{sstars}}
\tablehead{ID & $\alpha_{2000}$  & $\delta_{2000}$ & Type & P& 
$\theta$\tablenotemark{a} & Filter & Ref.  \\
&(hh:mm:ss.sss)&(dd:mm:ss.ss)&&(\%) &  ($\degr$) &&}
\startdata                    
   CMa R1 No. 24\tablenotemark{b} & 07:04:47.364 & -10:56:17.44 & Polarized & 2.1$\pm$ 0.05 & 86 $\pm$ 1& J & 1 \\     
   BD+28d4211 & 21:51:11.070  & +28:51:51.80  & Unpolarized & 0.041 $\pm$ 0.031 & 38.66 &    N\tablenotemark{c} & 2 \\
   & & & & 0.067 $\pm$ 0.023 & 135.00 & U & 2  \\
   & & & & 0.063 $\pm$ 0.023 & 30.30 & B & 2 \\
   & & & & 0.054 $\pm$ 0.027 & 54.22 & V & 2 \\
   G191B2B & 05:05:30.621 & +52:49:51.97 & Unpolarized & 0.065 $\pm$ 0.038 & 91.75 & U & 2 \\  
   & & & & 0.090 $\pm$ 0.048 & 156.82 & B & 2 \\ 
   & & & & 0.061 $\pm$ 0.038 & 147.65 & V & 2
\enddata
\tablenotetext{a}{Position angles measured from north to east.}
\tablenotetext{b}{
This is the only polarized standard star with well established polarization properties  in the $J$-band. 
}
\tablenotetext{c}{``Near-UV'' filter centered in 3450 \AA\ and with full width at 
half maximum (FWHM) bandpass of 650 \AA. For details, see  \citet{Schmidt92}.}
\tablerefs{(1)  \citet{Whittet92}; (2)  \citet{Schmidt92}, the two unpolarized stars are taken from a
 compilation of optical calibration data collected with the {\it Hubble Space Telescope}.}
\end{deluxetable}

\begin{deluxetable}{c c c c c c} 
\tablecaption{Observational results for the unpolarized standard stars\label{unpol}}
\tablewidth{0pt}
\tablehead{ID & Mission & P$_{obs}$  & P/${\sigma}$ & P$_{true}$ & P$_{min}$/P$_{max}$\tablenotemark{a} \\
& & (\%)  & & & (\%)}    
\startdata   
G191B2B &  2007 & 0.41  &    1.38  &  0.28  &  0.00/1.10  \\
BD+28d4211 & 2006 & 0.09  &   0.60  &  0.00  &  0.00/0.41  \\
BD+28d4211 & 2007 & 0.15  &   1.17  &  0.08  &  0.00/0.45 \\
\enddata
\tablenotetext{a}{Minimum and maximum value for the degree of polarization at 99\% 
confidence level \citep{Simmons85}.}
\end{deluxetable}

\begin{deluxetable}{c c c c c c c c} 
\tablecaption{Observational results for the polarized standard star\label{tab:pol}}
\tablewidth{0pt} 
\tablehead{ID & P$_{obs}$  & $\sigma_{P}$ & P/${\sigma}$ & P$_{true}$ & 
 P$_{min}$/P$_{max}$\tablenotemark{a} & $\theta_{obs}$ & $\sigma_{\theta}$ \\
 & (\%)  & (\%) & & (\%) & (\%) & (\degr) & (\degr)}  
\startdata  
CMa R1 &  2.10 & 0.26 &  7.98  &  2.08  &  1.38/ 2.78  &   92.7  & 4\\
 No 24 &  2.04  &  0.26 & 7.99  &  2.02  &  1.35/2.70  &   94.3  & 4 \\
 &  2.33  & 0.26 & 9.06  &  2.31  &  1.62/3.00  &   93.5  & 3  \\
 &  2.05  & 0.22 & 9.43  &  2.04  &  1.44/ 2.62  &   86.8  & 3 
\enddata
\tablenotetext{a}{Minimum and maximum value for the degree of polarization at 99\% 
confidence level \citep{Simmons85}.}  
\end{deluxetable}

\begin{deluxetable}{r c c c c c r c c c r r c c}
\tablecaption{Polarization data\label{tab:IRdata}}
\tabletypesize{\scriptsize}
\tablewidth{0pt}
\tablehead{ID & $\alpha_{2000}$ & $\delta_{2000}$ & $J$\tablenotemark{a} &  p$^R$ & 
$\sigma_{\mathrm{p^R}}$ & \multicolumn{1}{c}{$\theta^R$\tablenotemark{b}} &
p$^{J}$ & $\sigma_{\mathrm{p^J}}$ & p$^J_{\mathrm{true}}$ & \multicolumn{1}{c}{$\theta^J$\tablenotemark{b}} & $\sigma_{\theta}$\tablenotemark{c} & 
Rotator position & Class\tablenotemark{d} \\ 
& (hh:mm:ss.ss) & (dd:mm:ss.ss) & (mag) &  (\%) & (\%) & (\degr) & (\%) & (\%) & (\%) & (\degr) & (\degr) & (\degr) }
\startdata                    
1  & 03:29:14.58  & 31:06:38.20  &  14.70  & 4.04 & 0.75 & 151.5 & 1.74  & 1.27  & 1.41 & 173  & 27  &  0,90 & IR source \\
2  & 03:29:14.89  & 31:09:27.50  &  12.67  & 4.53 & 0.06 & 157.8 & 3.10  & 0.64  & 3.03 & 157  &  7  &  0,90 & Star  \\
3  & 03:29:15.05  & 31:08:06.90  &  16.06  & & & & 3.16  & 2.43  & 2.02 & 160  & 33  &  0,90 & IR source \\
4  & 03:29:16.08  & 31:07:31.40  &  12.99  & 4.51 & 0.12 & 166.2 & 2.32  & 0.74  & 2.20 & 172  & 10  &  0,90 & IR source \\
5  & 03:29:16.20  & 31:07:34.00  &  13.83  & 3.92 & 0.21 & 157.6 & 1.88  & 0.97  & 1.61 & 167  & 18  &  0,90 & --- \\
6  & 03:29:16.68  & 31:16:18.30  &  10.74  & 0.87 & 0.14 & 117.5 & 1.67  & 0.60  & 1.56 & 135  & 11  &  0    & IR source \\
7  & 03:29:17.52  & 31:07:33.20  &  15.04  & 3.88 & 0.25 & 163.3 & 1.82  & 1.45  & 1.10 & 171  & 37  &  0,90 & IR source \\
8  & 03:29:17.91  & 31:07:07.70  &  15.34  & & & & 2.70  & 1.91  & 1.90 & 169  & 29  &  0,90 & IR source \\
9  & 03:29:18.21  & 31:07:55.70  &  14.50  & 3.32 & 0.48 & 167.2 & 2.55  & 1.43  & 2.11 & 163  & 20  &  0,90 & IR source \\
10 & 03:29:18.64  & 31:09:59.60  &  12.50  & 4.40 & 0.04 & 156.4 & 4.68  & 0.65  & 4.63 & 155  &  4  &  0,90 & Star \\
11 & 03:29:20.01  & 31:09:54.30  &  12.45  & 5.75 & 0.10 & 164.0 & 4.89  & 0.64  & 4.85 & 160  &  4  &  0,90 & Star  \\
12 & 03:29:20.10  & 31:08:54.00  &  16.08  & & & & 3.15  & 2.44  & 1.99 & 155  & 33  &  0,90 & --- \\
13 & 03:29:21.87  & 31:15:36.30  &  11.33  & 0.94 & 0.09 & 49.4 & 1.60  & 0.75  & 1.41 &  49  & 16  &  0    & CTTS\tablenotemark{e} \\
14 & 03:29:23.50  & 31:07:25.00  &  16.57  & & & & 4.54  & 3.46  & 2.94 & 167  & 33  &  0,90 & --- \\
15 & 03:29:24.70  & 31.07:27.00  &  17.16  & & & & 4.56  & 4.42  & 1.12 & 167  & 24  &  0,90 & --- \\
16 & 03:29:25.60  & 31:08:43.00  &  17.00  & & & & 4.90  & 3.99  & 2.85 & 172  & 23  &  0,90 & IR source  \\
17 & 03:29:27.04  & 31:08:04.60  &  12.39  & 4.61 & 0.29 & 157.6 & 3.53  & 0.66  & 3.47 & 169  &  6  &  0,90 & IR source \\
18 & 03:29:27.16  & 31:06:48.20  &  14.92  & 5.20 & 0.51 & 166.1 & 3.32  & 1.79  & 2.80 & 173  & 20  &  0,90 & IR source \\
19 & 03:29:28.99  & 31:10:00.30  &  13.36  & & & & 4.01  & 0.93  & 3.90 & 142  &  7  &  0,90 & IR source  \\
20 & 03:29:29.60  & 31:08:47.90  &  16.30  & & & & 3.17  & 2.71  & 1.64 & 156  & 42  &  0,90 & IR source  \\
21 & 03:29:30.80  & 31.06:33.00  &  17.97  & & & & 7.01  & 6.51  & 2.61 & 160  & 25  &  0,90 & IR source  \\
22 & 03:29:32.41  & 31:13:01.10  &  13.34  & & & & 1.90  & 1.21  & 1.47 & 135  & 24  &  0    & IR source \\
\hline 
\noalign{\vskip .8ex}
\multicolumn{14}{c}{stars with $R$-band data only} \\
\noalign{\vskip .8ex}
\hline 
\noalign{\vskip .8ex}
23 & 03:29:02.87 & 31:16:00.82  & 12.84 &  0.68 &   0.15  &   66.4 & & & & & & & YSOC \\
24 & 03:29:03.74 & 31:16:03.60  & 11.74 &  7.58 &   0.54  &   57.7 & & & & & & & V512 Per \\
25 & 03:29:04.04 & 31:17:06.66  &  13.31 &  1.38 &   0.52  &   86.3 & & & & & & & BD\tablenotemark{f} \\
26 & 03:29:07.39 & 31:10:49.02  &  13.10 &  3.11 &   0.17  &  167.0 & & & & & & & Star \\
27 & 03:29:09.57 & 31:09:08.68  &  14.94 &  4.61 &   0.61  &  161.1 & & & & & & & Star \\
28 & 03:29:12.16 & 31:08:10.91  &  12.99 &  0.39 &   0.08  &   71.8 & & & & & & & IR source \\
29 & 03:29:17.84 & 31:05:37.40  &  14.04 &  4.35 &   0.45  &  164.4 & & & & & & & IR source \\
30 & 03:29:20.59 & 31:06:11.48  &  15.28 &  5.13 &   0.93  &  174.5 & & & & & & & IR source \\
31 & 03:29:29.11 & 31:06:08.77  &  14.07 &  5.36 &   0.32  &  167.3 & & & & & & & IR source \\
32 & 03:29:34.24 & 31:07:53.33  &  13.64 &  2.15 &   0.39  &  153.7 & & & & & & & IR source \\
33 & 03:29:39.77 & 31:14:51.61  &  &  1.38 &   0.19  &   13.6 & & & & & & & --- \\
34 & 03:29:40.43 & 31:12:46.38  &  13.04 &  0.34 &   0.06  &   41.4 & & & & & & & IR source
\enddata
\tablenotetext{a}{
LIRIS $J$ magnitude for stars observed with this camera, 2MASS $J$ magnitude for stars observed only in the $R$-band.}
\tablenotetext{b}{Position angles are counted from north to east.}
\tablenotetext{c}{1$\sigma$ uncertainty of the position angle (see text for explanation on how it was estimated).}
\tablenotetext{d}{Object's type as found in the SIMBAD Astronomical Database.}
\tablenotetext{e}{Classical T-Tauri Star \citep{Lada74}.}
\tablenotetext{f}{
Brown Dwarf \citep{IGO10}.}
\end{deluxetable} 
        
\end{document}